\documentclass[sigconf]{acmart}
\AtBeginDocument{%
  \providecommand\BibTeX{{%
    \normalfont B\kern-0.5em{\scshape i\kern-0.25em b}\kern-0.8em\TeX}}}

\setcopyright{acmcopyright}
\copyrightyear{2018}
\acmYear{2018}
\acmDOI{XXXXXXX.XXXXXXX}

\acmConference[Conference acronym 'XX]{Make sure to enter the correct
  conference title from your rights confirmation emai}{June 03--05,
  2018}{Woodstock, NY}
\acmPrice{15.00}
\acmISBN{978-1-4503-XXXX-X/18/06}

\usepackage{rotating}

\begin{document}

\title{Robust Quantification of Gender Disparity in Pre-Modern English Literature using Natural Language Processing}

 \author{Akarsh Nagaraj}
 \authornotemark[1]
 \email{akarshna@usc.edu}
  \author{Mayank Kejriwal}
 \authornote{Both authors contributed equally to this research.}
 \email{kejriwal@isi.edu}
 \orcid{1234-5678-9012}
 \affiliation{%
   \institution{Information Sciences Institute \\ University of Southern California}
   \streetaddress{4676 Admiralty Way}
   \city{Marina del Rey}
   \state{California}
   \country{USA}
   \postcode{90292}
 }

\renewcommand{\shortauthors}{Anonymous}

\begin{abstract}
  Research has continued to shed light on the extent and significance of gender disparity in social, cultural and economic spheres. More recently, computational tools from the Natural Language Processing (NLP) literature have been proposed for measuring such disparity using relatively extensive datasets and empirically rigorous methodologies. In this paper, we contribute to this line of research by studying gender disparity, at scale, in copyright-expired literary texts published in the pre-modern period (defined in this work as the period ranging from the mid-nineteenth through the mid-twentieth century). One of the challenges in using such tools is to ensure quality control, and by extension, trustworthy statistical analysis. Another challenge is in using materials and methods that are publicly available and have been established for some time, both to ensure that they can be used and vetted in the future, and also, to add confidence to the methodology itself. We present our solution to addressing these challenges, and using multiple measures, demonstrate the significant discrepancy between the prevalence of female characters and male characters in pre-modern literature. The evidence suggests that the discrepancy declines when the author is female. The discrepancy seems to be relatively stable as we plot data over the decades in this century-long period. Finally, we aim to carefully describe both the limitations and ethical caveats associated with this study, and others like it. 
\end{abstract}




\maketitle

\section{Introduction}
Recent innovations in deep neural networks have led to impressive advances in Natural Language Processing (NLP)  \cite{bert, transformers}. These advances include new state-of-the-art results in tasks as diverse as question answering, information extraction, sentiment analysis, conversational `chatbot' agents and summarization, to only name a few \cite{qa1, transformerIE, transformerSA, siblini2019multilingual, transformerSum}. Due to the performance of these models, it has also become possible in recent years to use NLP tools for computational social science and digital humanities \cite{NLPcomp1, NLPcomp2, NLPcomp3, NLPcomp4} (see also \emph{Related Work}). Such methods are especially important for obtaining quantitative results at scale on large datasets that are not possible to examine in a fully manual manner without expending extensive labor and cost. 

In this article, we use computational methods from the NLP community, implemented in open-source, industrial-grade packages, to quantify and explore the phenomenon of \emph{gender-specific character prevalence} in pre-modern literature. Gender-specific character prevalence intuitively measures the extent to which a reader is likely to encounter female characters, compared to male characters, in a given book. For our empirical analysis, we use the publicly available English-language books in the Project Gutenberg corpus (with links provided subsequently in \emph{Materials and Methods}), which contains copyright-expired, pre-modern books that can be used for such studies. 

As noted below, there are several methodological avenues for defining and measuring character prevalence. Therefore, to ensure that our results are robust, we consider three measurements of gender-specific character prevalence, along with detailed statistical analysis. We also consider whether differences in gender-specific character prevalence have declined over time, and if controlling for the gender of a book's author yields statistically different results. 

Specifically, we investigate three related hypotheses in this work:

{\bf Hypothesis 1: Female character prevalence is less than that of male characters.} We consider three different prevalence measures to achieve robust findings. For example, while one of our measures simply counts the numbers of male and female characters extracted from the book-text, other measures count the number of occurrences of such characters, and also take the occurrences of male and female pronouns into account. More details are provided in a subsequent section where we discuss the experimental methodology.

{\bf Hypothesis 2: The difference between male and female character prevalence significantly declines  when controlling for authors' gender.} We conduct similar experiments as for Hypothesis 1, but we control for the authorship of the book to understand whether the gender of the author has an impact on the conclusions when used as a control.  

{\bf Hypothesis 3: Female character prevalence changes significantly between 1800 to 1950.} We plot character prevalence over time, based on the most active year of the author of the books, to understand if female character prevalence has  been increasing relative to male character exposure (at least approximately) over time. 
An important limitation of this study that we note at the outset, and that is described and contextualized in much more detail in Section \ref{ethics} is that we only study the disparity between male and female characters. Unfortunately, the disparity of non-binary and trans characters compared to traditional genders could not be accurately studied due to a lack of computational tools for extracting characters whose genders do not fall in the dichotomous categories of male and female. We believe that this highlights a pressing need to develop such tools, an issue that we comment more on, in Section \ref{ethics}. 

The rest of this article is structured as follows. We begin with a brief review of the \emph{Related Work} in this area, followed by details on the \emph{Materials and Methods} used in this paper for investigating the three hypotheses stated earlier. Specifically, we detail not only the dataset, but also the NLP pipeline that we constructed using open-source software for conducting this study. Our code and data are open-source and replicable. We describe our key findings in \emph{Results}, followed by a more qualitative assessment in \emph{Discussion}. Importantly, before concluding the paper, we enumerate, based on our understanding and analysis, of the limitations of the study, relevant ethical issues and some guidance to future researchers on addressing them.

\section{Related Work}

A number of recent social science studies, some of which are computational, have shown that gender bias continues to exist in many aspects of economic, social and cultural life, including movies \cite{bias4}, executive positions in top corporations \cite{bias1}, board membership \cite{bias2}, and political leadership \cite{bias3}. While researchers have tackled the challenge of accurately measuring gender bias in large corpora like Wikipedia  using computational methodologies \cite{reagle2011gender}, such studies have been generally lacking in literature and cultural corpora, especially those that have not been published first on the Web (and thereby lack additional context, such as hyperlinks). Until quite recently, using NLP techniques for such studies was problematic because of concerns over quality, given the importance of this issue. However, the improvements in NLP cited earlier suggest that the time is ripe for conducting such computational social science studies, with appropriate quality control measures in place \cite{NLPcomp3}. 

More recently, textual analysis of novels and literary texts, especially using structural and statistical analyses, has been a major theme in Digital Humanities research. We cite the book by Jockers on digital methods and literary history as a notable example in this regard \cite{jockers2013macroanalysis}. Other work by Jockers, done with co-authors, is also relevant, including his study (with Mimno) on literary themes in novels and other 19th century texts \cite{jockers2013significant}, and his study (with Archer) of the structure of bestselling novels \cite{archer2016bestseller}, among others \cite{lambert2020pace, jockers2014text}. 

The \emph{Gender in Novels} project is an example that closely matches the goals of this work and was conducted independently\footnote{\url{http://gendernovels.digitalhumanitiesmit.org/info/gender_novels_overview}}, within the MIT Digital Humanities Lab. Similar to the research herein, their goal is to ``study the ever-adapting and changing view on gender by writers all around the globe in the nineteenth and twentieth centuries.'' Other computational studies on gender, some of which involve multi-modal data such as images and video, include \cite{compgender1, compgender2, compgender3, compgender4, compgender5}. 

There is also a long line of gender studies literature in the humanities that is complementary to the research herein \cite{genderhum1, genderhum2, genderhum3}; however, many of those articles use qualitative methods and tend to deeply analyze relatively small corpora, such as the books of a single author \cite{homans1993dinah}, or within a narrow period or genre \cite{fine1998gender}, rather than a broad-based computational analysis such as in this work or other similar projects such as \emph{Gender in Novels}. To take just example, the work in \cite{homans1993dinah} studies both gender and class issues in George Eliot's early novels. Other excellent and general examples of gender studies in the humanities include \cite{genhum1, genhum2, genhum3}. These works are necessary for understanding and contextualizing gender in literature and the humanities, and this context should be borne in mind for a causal interpretation of the statistical and computational results that we present in this paper.


\section{Materials and Methods}

\subsection{Data}
The raw data for the study was obtained from Project Gutenberg\footnote{\url{https://www.gutenberg.org/}}, and originally comprised 3,036 English books represented as text files, penned by 142 authors (of whom 14 are female) between 1700 and 1950. We used a subset of the 3,036 books for this study by selecting works that are fictional, and have been published between 1800 and 1950 (inclusive). Following this filtering, the final dataset comprised 2,426 books that cover fictional genres ranging from adventure and science fiction, to mystery and romance. Not all works are necessarily in the format of a novel, since the corpus also includes short stories, plays and poems.

\subsection{Data Preprocessing, Character Extraction and Gender Classification}

An important contribution of this study is the quantification of gender-specific character prevalence in literature within a sufficiently broad corpus. Since manually extracting characters and character occurrences from a corpus of 2,426 books is not feasible, we propose to use high-performance NLP methods to extract characters in various ways, and to automatically classify whether they are male or female. Once extracted, and tagged with gender, robust analysis of character exposure becomes feasible.   

The primary NLP method that we rely on to extract characters is Named Entity Recognition (NER) \cite{NER1}, which also goes by \emph{named entity identification, entity chunking,} and \emph{entity extraction} in the literature \cite{NEI,EC,EE}, is a specific type of the broader \emph{information extraction} (IE) problem that has witnessed considerable advances in recent years, including for domain-specific corpora \cite{transformerIE, IE1, IE2, IE3}. NER seeks to locate and classify named entities mentioned in unstructured text into pre-defined categories such as person names, organizations, locations, monetary values, to name a few. In our study, we are interested in extracting person names from the text of the books. 

In order to apply NER, we first need to split the input text into sentences. Identifying sentences is important because they form logical units of thought and represent the borders of many grammatical effects. To do so, we employ \emph{sentence segmentation} as the first pre-processing step. Sentence segmentation is an important early step in many NLP pipelines \cite{sentenceSeg}. For example, character extraction (an instance of NER) algorithms tend to be more accurate and efficient when they are executed on shorter self-contained spans of text (such as sentences) rather than on entire corpora.  Despite its seeming simplicity, a generalizable implementation of sentence segmentation is non-trivial for various language-specific reasons. One example is the use of periods in abbreviations and numbers, in addition to its more common use at the end of sentences. 

Recent NLP software packages have achieved impressive results in a variety of tasks, including sentence segmentation, primarily due to the advent of deep learning and maturity of `language representation' models. For our purposes, we used the sentence segmentation module from a Python library called SegTok\footnote{\url{http://fnl.es/segtok-a-segmentation-and-tokenization-library.html}}, developed to process orthographically regular Germanic languages, of which English is an example. SegTok is capable of identifying sentence terminals such as `.', `?' and `!' and \emph{disambiguating} them when they appear in the middle of a sentence (like in the case of abbreviations and website links), which significantly reduces the probability that a sentence is segmented before it has truly concluded. After executing SegTok on each book in our corpus, we also manually assessed its performance by sampling `challenging' sentences (that contained inconsistent sentence terminals, including periods in the middle of the sentence, as in the cases noted above) and verifying that the full sentence was correctly segmented by the software. 

Before deciding on SegTok, we also tried other viable sentence segmentation packages in the NLP literature, such as \emph{PunktSentenceTokenizer}\footnote{\url{https://www.nltk.org/_modules/nltk/tokenize/punkt.html}}, which is part of the Natural Language Tool Kit (NLTK) \cite{nltk}. NLTK is a well-known suite of libraries and packages for symbolic and statistical natural language tasks like text classification, tokenization, stemming, tagging, and parsing. We found that PunktSentenceTokenizer was incorrectly splitting sentences by abbreviation-periods in some of our sample texts. Furthermore, SegTok also had faster execution times, which is essential when processing large texts like the Project Gutenberg corpus of books.
Additionally, we evaluated the accuracy of SegTok by randomly sampling 110 sentence outputs that were segmented, and manually tagging them as being correctly segmented with respect to the paragraph in which the sentence was originally embedded. We found that, of these 110 sentences, only two were incorrectly segmented,\footnote{We found that, of the two sentences that were incorrectly segmented, the error was minor. In both cases, there were words in all caps either at the beginning or end of the sentence. This may have been due to (for example) a chapter header, or slight formatting discrepancies in the underlying corpus itself. In either case, the `correct' segmentation was always a subset of the actual output.} yielding an accuracy of 98.18\%. Hence, SegTok was chosen for segmenting the text of each of the books in our corpus into sentences.

\subsubsection{Character Extraction, Disambiguation and Gender Classification}
In order to measure gender-specific character prevalence, which is required for all three of our hypotheses, we need to count the numbers of male and female characters in each of the books in the corpus. As noted earlier, extracting person names from the text is an instance of the NER problem. Similar to the methodology for selecting a viable sentence segmentation module, we compared the performance of two popular NER libraries in the NLP community - \emph{SpaCy}\footnote{\url{https://spacy.io/api}} \cite{spacy}, an industrial-scale open-source software library for advanced NLP, typically relying on neural network models for part-of-speech tagging, dependency parsing, text categorization and NER; and \emph{NE\_Chunk}\footnote{\url{https://www.nltk.org/api/nltk.chunk.html}}, which is also part of NLTK. Named entity chunking extracts `chunks' (akin to phrases) from sentences and assigns them semantic tags, such as person, location and organization. We found that NE\_Chunk consistently outperformed SpaCy by achieving near 100\% precision in extracting characters (`person' entities) from a sample set of books\footnote{Specifically, we randomly chose four books and manually identified a total of 72 main characters in those books. We then compared these 72 `gold standard' characters to the ones extracted by NE\_Chunk. We found that only one (male) character was not correctly extracted.}. Hence, we chose it as the character extraction tool.

However, since characters are independently extracted from each sentence, multiple occurrences of the same character can be extracted across all sentences in a book due to artifacts such as slightly different spelling usage or use of only first or last names, instead of the full name. To discover different occurrences (of the same character), we used a Python library called \emph{difflib}\footnote{\url{https://docs.python.org/3/library/difflib.html}}, which provides classes and functions for comparing sequences. Specifically, the \emph{SequenceMatcher} class from difflib library compares two strings and provides a similarity score between 0 (no match at all) to 1 (complete match, i.e., strings are the same). We used the SequenceMatcher class to perform disambiguation of the characters and link the different versions of the same character together. Specifically, string pairs with a similarity score of 0.7 or above were treated as duplicates. This threshold was selected after some sampling and manual verification. This deduplication allows us to count the number of \emph{unique} characters extracted from the text of each book. To assess its accuracy, we randomly sampled 76 character pairs that were disambiguated as duplicates by this heuristic technique, and found that 72 were correctly disambiguated, yielding an accuracy of 94.74\%. The errors in disambiguation primarily arose from false positives and mostly involved the names of monarchs e.g., George III and George IV were incorrectly identified as duplicates due to high string similarity.  

Finally, to classify the extracted characters as male and female, we used \emph{Gender\_Detector}\footnote{\url{https://pypi.org/project/gender-detector/}}, a Python library developed using data from the \emph{Global Name Data} project\footnote{\url{https://github.com/OpenGenderTracking/globalnamedata}}, which is able to determine the gender of a character from the first name. Using this library, we were able to heuristically tag each extracted character as male or female. We evaluated the accuracy of this method by randomly sampling 100 extracted characters\footnote{50 male, and 50 female, to avoid gender-specific skewness in computing accuracy estimates.} and manually checking their actual gender against the predicted gender. There was only one error,\footnote{The single error was due to a female character named `Captain Leslie' being erroneously tagged as male.} yielding an accuracy of 99\%.

\subsection{Availability of Processed Data for Replication}

Although only publicly available packages have been used in the NLP-based pipeline described earlier, we recognize that the steps noted above are non-trivial to execute, especially by social scientists and digital humanities scholars who may want to use the data to run their own studies, or replicate the results described subsequently. Hence, we have published the processed version of the data as a peer-reviewed data-in-brief article, with links to a Mendeley repository containing the processed data\footnote{Due to author anonymity requirements, we withhold the citation in this version for the repository and processed dataset.}. 

\subsection{Experimental Methodology}
In this section, we present three definitions, and corresponding measurements, of character prevalence for male and female genders:
\begin{enumerate}
    \item {\bf Character Count}: The numbers of unique male and female characters extracted from each of the books.
    \item {\bf  Character Occurrence Count}: The number of times (or `occurrences') male and female characters are mentioned by name in the book.
    \item {\bf Pronoun Count}: The numbers of male pronouns (he, him, his) and female pronouns (she, her, hers) present in each book. We do not address gender-neutral pronouns in this article as we could not identify, despite our best efforts, a publicly available or established computational package for distinguishing the uses of pronouns, such as `they', as referring to multiple individuals or a single individual. However, gender-neutral pronouns referring to individuals that choose their pronouns as `they', `their' or `them' could be considered in future research that draws on more modern books to replicate the following analyses.
\end{enumerate}

To investigate Hypothesis 2, we clearly need data on authors' genders. Since there are 142 unique authors in our corpus, we manually tagged their gender using both their names, and public resources like Wikipedia, and found only 14 to be female\footnote{We recognize the limitations of this approach, and contextualize it further in Section \ref{ethics}.}. Finally, to investigate Hypothesis 3, we need to plot a time-series of gender-specific character prevalence. Unfortunately, the books in the Project Gutenberg corpus do not contain metadata that indicates the date of publication of an individual book. To be both robust and fair, while obtaining the findings of interest for the purposes of investigating Hypothesis 3, we made two approximations. First, we assigned a year to a book by considering that book's author's most active year (defined as the year in which the author published the most books). All books by that author were tagged with that year. Second, to account for the noise that this may cause, we compute rolling averages of character-prevalence over ten-year windows at a time. 

\section{Results}
In this section, we describe our key findings. For ease of exposition, we present the findings pertinent to each hypothesis in turn.

\subsection{Hypothesis 1}
\begin{figure*}
\includegraphics[width=\textwidth]{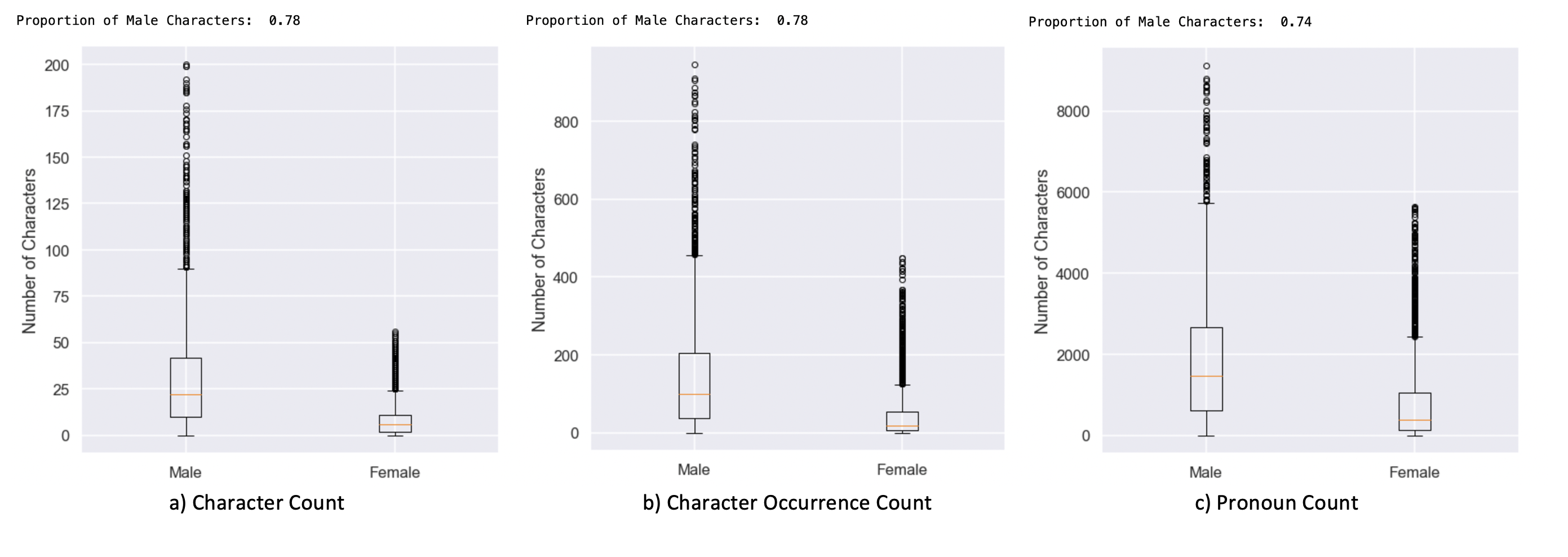}
\caption{Summary statistics (illustrated as boxplots) characterizing differences between male and female character exposure using the three exposure measures defined in the \emph{Experimental Methodology} subsection. The orange line represents the median, with circles indicating outliers. \emph{Proportion of Male Characters} for a given measure is the average (over all books) of the ratio of the male character count in a book to the total character count in that book. } 
\label{fig1}
\end{figure*}

In the \emph{Introduction}, we stated Hypothesis 1 as the claim that female character prevalence is less than male character prevalence. In the previous section, we also presented three different ways in which gender-specific character prevalence can be computed. Using these three different measures, we plot in Fig \ref{fig1} the differences between male and female character prevalence. 

The figure yields some important insights, the most important being that, no matter the measure used for quantifying character prevalence, female character prevalence in the books in our corpus is significantly lower than male character prevalence. While the magnitude of the difference is striking in all cases, it does depend on the specific measure employed. The largest difference, on average, is observed when using the \emph{Character Count} measure (78\% for male character prevalence), while the smallest difference is observed when using the \emph{Pronoun Count} measure (74\% for male character prevalence). In testing for significance by using the one-sided independent Student's t-test, we find that the difference between the means of male and female measures is highly significant with a reported p-value (for each of the three measures) close to zero ($<< 10^{-100}$). 

In considering absolute counts, we find a mean of 32 unique male characters per book, compared to only 9 unique female characters per book. The result does not significantly change even when accounting for outliers by using the median instead of the mean. The median numbers of unique male and female characters are 22 and 6, respectively. Taken collectively, the findings robustly bear out the claim in Hypothesis 1 that character prevalence for female characters is significantly lower than that for male characters.   

\subsection{Hypothesis 2}
\begin{figure*}
\includegraphics[height=7.8in]{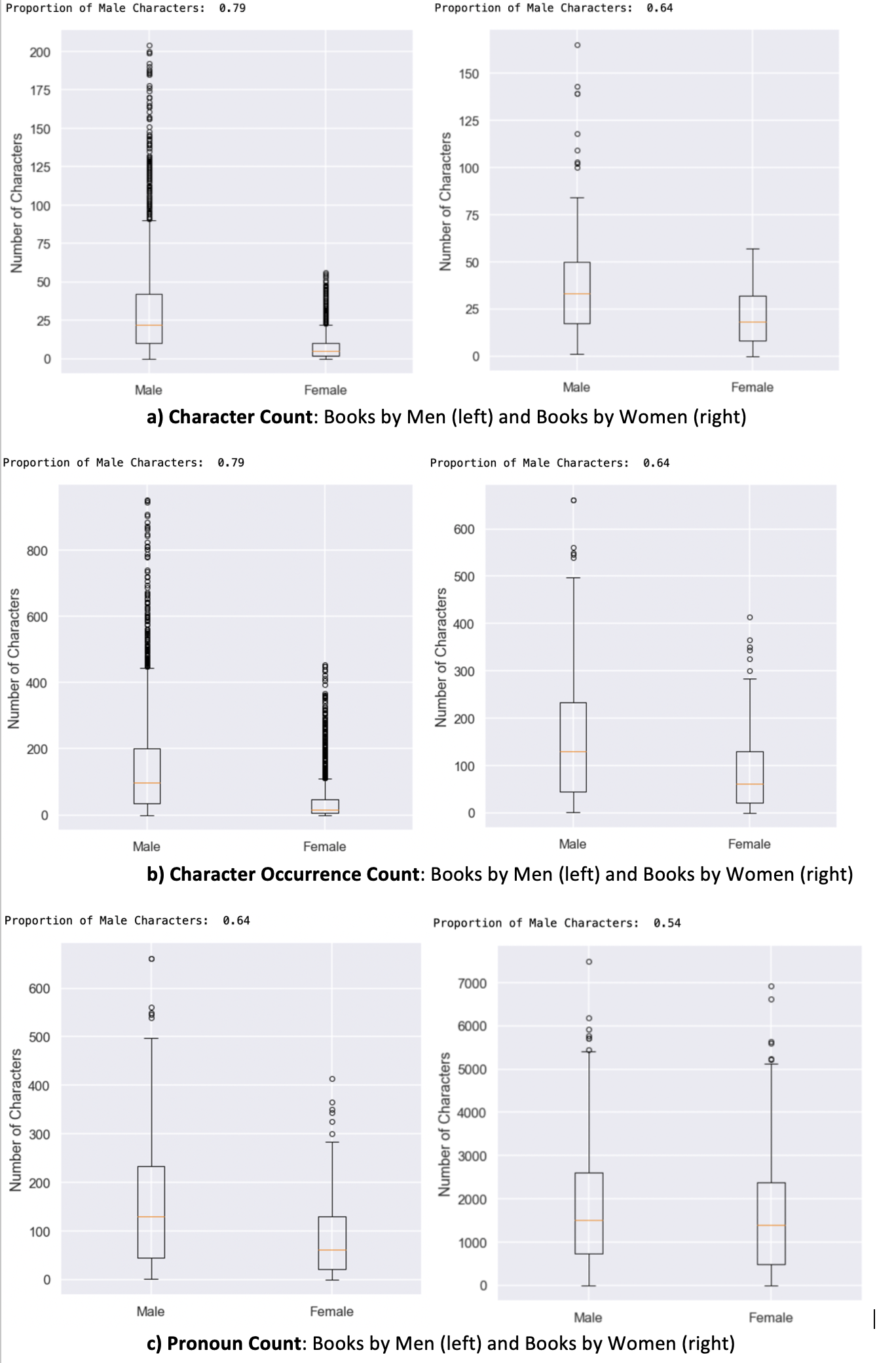}
\caption{Summary statistics (illustrated as boxplots) characterizing differences between male and female character exposure using the three exposure measures with gender of the author as control. \emph{Proportion of Male Characters} for a given measure is the average (over all books) of the ratio of the male character count in a book to the total character count in that book.} 
\label{fig2}
\end{figure*}

Hypothesis 2 was the claim that the difference between male and female character prevalence, found to be significant and large in the previous investigation, could decline if we control for the gender of the authors. As shown in Fig \ref{fig2}, we again use the three character prevalence measures defined earlier to investigate this hypothesis. The figure shows that the proportion of male characters falls from 79\% in male-authored books to 64\% in female-authored books when using \emph{Character Count} measure. On average, there are 32 (unique) male and 8 female characters per male-authored book compared to 38 male and 21 female characters in female-authored books.

Similarly, when using the \emph{Character Occurrence Count} measure instead, the proportion of male character occurrences drops from 79\% in male-authored books to 64\% in female-authored books. The third measure of gender prevalence, \emph{Pronoun Count}, concurs with the other two measures, in that the proportion of male pronouns shows a decline from 64\% in male-authored books to 54\% in female-authored books.

In summary, there is good empirical support in favor of Hypothesis 2. Based on all three gender-specific character prevalence measures, controlling for the gender of the author shows that the differences noted in the Hypothesis 1 results are diminished, with  female characters being significantly more prevalent in female-authored books, compared to male-authored books.

\subsection{Hypothesis 3}

\begin{figure*}
\includegraphics[width=\textwidth]{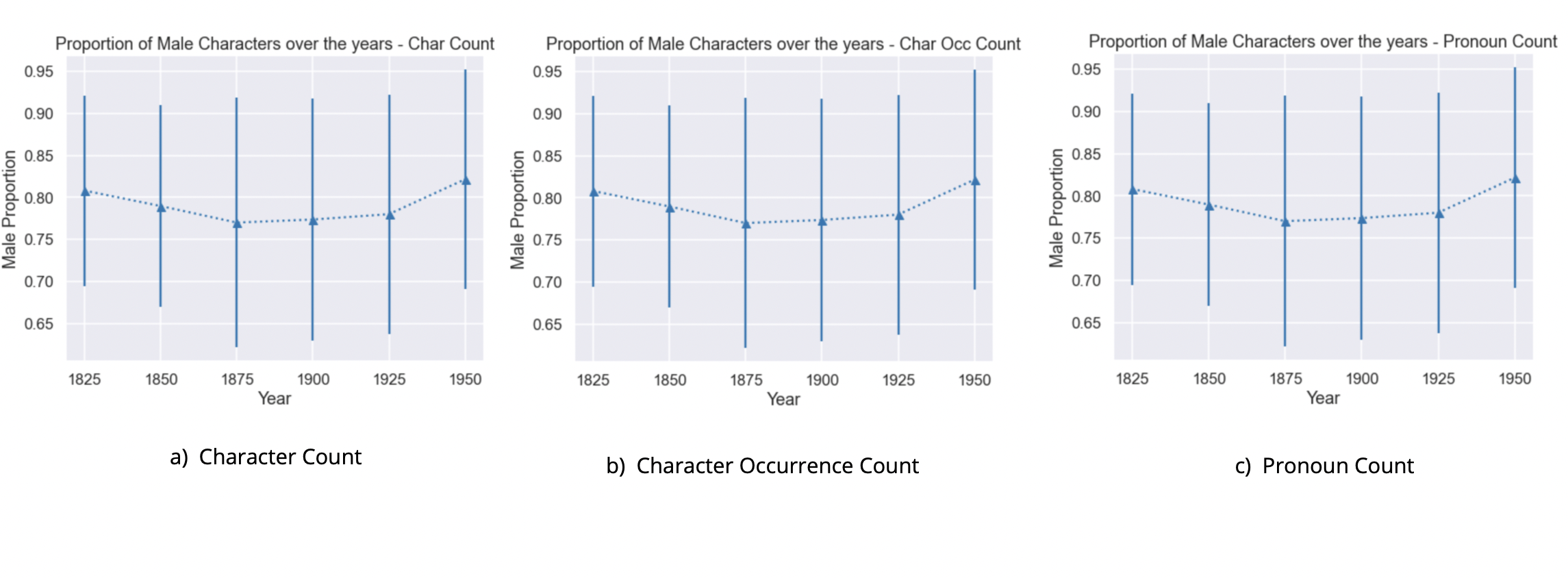}
\caption{Error Bar Plots illustrating the change in proportion of male characters using the three prevalence measures defined in the \emph{Experimental Methodology} subsection. The horizontal axis denotes the year (split into buckets) the book was published while vertical axis denotes the ratio of male characters to total characters. The triangle depicts the mean proportion of male characters while the vertical error bars show the 95\% confidence interval.} 
\label{fig3}
\end{figure*}

Finally, Hypothesis 3 was the claim that female character prevalence changes significantly between 1800 to 1950. Fig \ref{fig3} demonstrates the results by plotting the male character prevalence (using the three different measures) over the years. The methodology for determining the year was described earlier in \emph{Experimental Methodology}. Unlike the other two hypothesis, we do not see empirical support in Fig \ref{fig3} for Hypothesis 3. In fact, the male proportion of characters does not seem to change significantly over the years, with all three character prevalence measures showing that the mean (of male character exposure) for books penned between 1800 to 1950 varies only in the range of 75\%-80\%. Even over a long period, the female character prevalence in the corpus does not change much.

These results are consistent with some related findings in modern film. For example, as shown by some recent authors \cite{yang2020measuring}, until very recently, meaningful representation of women in movies has been low using a number of metrics. Even in recent data, a discrepancy remains. An equivalent literature-based survey is not feasible at this time, since the text files of the books published in the last few decades (mostly after the post-war period) are not in the public domain and available for study, unlike the books in the Project Gutenberg corpus.

\section{Discussion}

Our experiments indicate the severe imbalance of gender prevalence in the sample (available in the Project Gutenberg corpus) of books written during the period between 1800 and 1950. In particular, the experiments illustrate robust support for both Hypothesis 1 and 2. The former is supported by the fact that, when gender prevalence is compared using each of the three measures, on average, 3 out of 4 characters (or character occurrences, including using pronouns) in the books are found to be male. The latter is supported by the fact that these gender-specific differences diminish when we consider female-authored books to male-authored books.


Even worse, explorations concerning Hypothesis 3 revealed that the female character prevalence did not change much over the years from 1800 to 1950. Recent studies on gender equality have showed that over the last few decades, female characters' exposure in films has increased \cite{yang2020measuring}, but is still lower than that of male characters. They also found that, in movies involving female screenwriters and filmmakers, female representation and character development on screen is considerably higher. We expected to see similar and steady improvements in the long period between 1800 and 1950, but experimental results on our sample of books do not bear this out. At least over the period in question, female representation in fictional literature seems to have stayed largely stagnant. Representation of female authors is also low compared to male authors. 

While our experiments illustrated the quantitative differences in gender prevalence, in this section, we also attempt a qualitative assessment by systematically analyzing the kinds of words associated with male and female character occurrences, using computational techniques from NLP. To do so, we first extracted 5 sentences  around the first occurrence of each character (specifically, 1 sentence before and 4 after), which is where the description and introduction of the character is generally present. Then, we filtered all the words in these sentences and retained only the adjectives. We used part-of-speech (POS) tagging to accomplish this step \cite{postagging}, which is a process of converting a sentence to list of tuples, where each tuple comprises a word in the sentence, with a corresponding tag indicating whether the word is a noun, adjective, verb, and so on. Using POS tagging, we only retained words that were adjectives. In extracting adjectives around the first occurrence of each of the characters, our intent was to understand the descriptive theme and topics associated with male and female characters when they are first introduced in the book. 

To find these topics and themes, we relied on an NLP technique called \emph{word embeddings} \cite{word2vec}, where a neural network is used to `embed' each word in a corpus as a continuous, real-valued vector with a few hundred dimensions. The original neural network-based systems for word embeddings, such as word2vec, operated by sliding a window of a pre-specified size over the sequences of words in the corpus, and optimizing an objective function, such that vectors of words that tend to occur in the same window frequently are `close' to each other in the embedding space. Empirically, it was found that words that have similar meaning and semantic relations tended to be embedded closer together when a sufficiently large and representative corpus of text is used (such as Wikipedia, or the Google News corpus). As further validation of these embeddings, a number of operations, including analogies, are found to hold naturally in the vector space. For example, the resulting vector for $\vec{King}-\vec{Man}+\vec{Woman}$ was found to lie very close to $\vec{Queen}$ \cite{word2vec}. With the advent of transformer neural networks \cite{transformers}, language representation learning has become more complex and context-sensitive \cite{bert,unifiedqa}, although for this preliminary experiment we only consider a robust version of the word2vec model, described and linked below.   

An advantage of words being represented as vectors, capturing some notion of natural semantics in the embedding space, is that we can use the vector representations of the adjectives  within an unsupervised \emph{clustering} framework to recover themes and topics. We used the publicly available pre-trained Wiki News word embeddings that were derived by executing the popular fastText package on a large corpus of text\footnote{\url{https://dl.fbaipublicfiles.com/fasttext/vectors-english/wiki-news-300d-1M.vec.zip}}. The fastText package can be used for learning of word embeddings while being robust to misspellings and minor variations, and was originally created and released by Facebook's AI Research lab \cite{fasttext}. This model can be used to embed words in new text into vectors that capture semantic properties of words in a continuous-dimension space, as described above. 

Specifically, we obtained two sets of vectors, one each for male and female, respectively containing the embeddings of adjectives extracted around male and female characters. Next, we clustered these words using the classic k-Means algorithm into 8 clusters. The number of clusters was chosen as 8, since it was found to provide meaningfully different clusters without much overlap. We initially started with a smaller value for $k$ but incrementally increased it until qualitatively meaningful clusters were visible. In future work, one could automatically set the value for $k$ by using one of many heuristic methods, including the elbow method  \cite{elbow}.   



\begin{figure}
\includegraphics[height=6.8in]{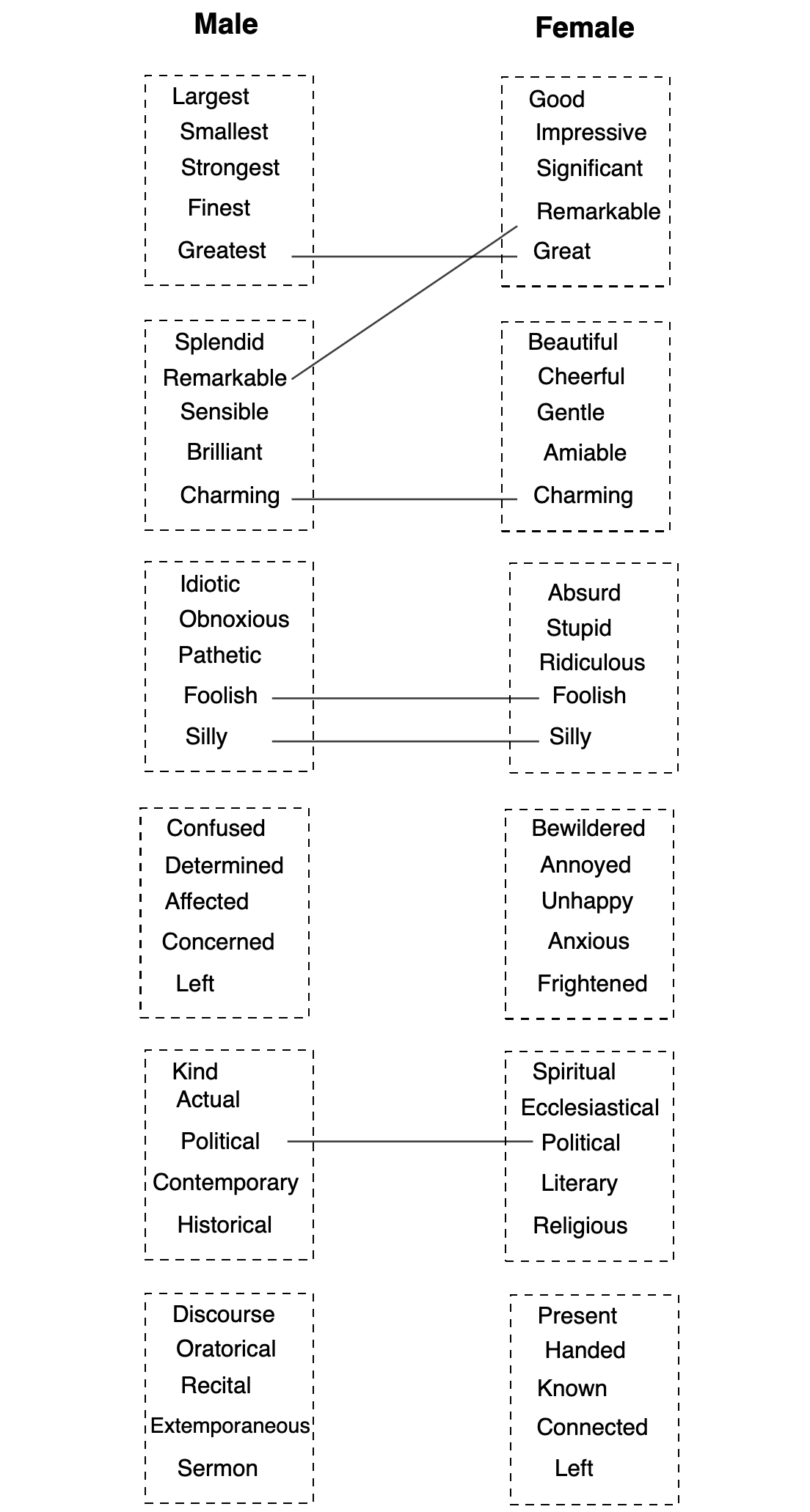}
\caption{A comparison of the most representative words in six word-embedding clusters for both male and female characters.} 
\label{fig4}
\end{figure}

Once the clusters were obtained, we took the `mid-point' or centroid of each cluster, and recovered the 5 words nearest to the centroid in the vector space. In this manner, we use these 5 words (per cluster) to approximately represent the main theme of that cluster. For ease of visualization, the results of 6 out of the 8 clusters are illustrated in Fig \ref{fig4}. The remaining two clusters covered themes such as nationality (e.g., `British', `American'), and were excluded from the visualization as they were largely similar between the two genders. 

Although there are some similarities between the representative words across genders, we also found that while male-adjectives clusters contain words like `strongest', `largest', `obnoxious', and `sensible', female-adjectives clusters tended to contain words like `beautiful', `amiable', `gentle' and `frightened'.

At the same time, it is worth noting that this method is qualitative and heuristic, and that more experiments are needed to understand differences in descriptions of male versus female characters. Furthermore, there may also be overlap between words when male and characters are introduced in the same paragraph. However, the experiment also helps us to understand differences in themes when describing male versus female characters. By using a similar methodology, other questions may also be explored, including equivalent versions of Hypotheses 2 and 3. For example, a version of the methodology could be used to explore the question of whether descriptions have changes over time, or are different between cross-section samples of male-authored versus female-authored books.

\section{Limitations of Study and Ethical Issues}\label{ethics} 

Gender, and gender identity, are important and complex issues in society, on which our understanding continues to evolve. In light of this complexity, it is important to highlight both the limitations of this study, and the limitations of our findings therein, as well as ethical caveats that future researchers must bear in mind when interpreting our findings, or even using the data and methods described in this work to obtain their own findings. 

First, and perhaps most importantly, we openly acknowledge a fundamental limitation of this study as one that only considered a dichotomous male-female gender categorization. Unfortunately, despite the best of our efforts, we did not find methods in the NLP literature that would allow us to detect non-binary, non-conforming and transgender individuals with the necessary accuracy. The \emph{Gender\_Detector} package that was previously described does not offer such a capability. Accuracy is paramount because non-conforming genders have already faced high levels of oppression historically, and it is not evident that they have been conveyed in literature as directly or representative of the populace as male characters. Considering the large disparity we already witness in our findings for female characters, we also cannot rule out complete suppression of (traditional and dichotomous) gender non-conformity. Indeed, we hope that future studies will make direct use of our data to study this issue in depth, in the same way that this study sought to convey the high levels of female character under-representation in pre-modern English literature. 

Second, our study is obviously confined to the subset of books that we considered. While the set we did consider withstood the test of time among the books in that period, the population in that period was exposed to a broader set of literature (including books, pamphlets, plays and so on), which may yield different statistics compared to this study. However, as we showed in the related work, our estimates of female character under-representation agree to some extent with recent statistics on female character representation on screen, or in scenes with meaningful dialogue.

The other limitations noted here are related to the processing of the dataset. 
We summarize them below:

\begin{enumerate}
    \item {\bf Gender from Name Assumption:} An assumption made by our pipeline was that gender could be determined from the names of book authors or characters. Although we made some effort to verify the gender from other sources of information, such as Wikipedia articles, and have a near-perfect accuracy estimate based on sampled manual annotations, we could not do so for all authors and characters. There may be some bias of which we may not be fully aware. Certainly, we caution other scholars on solely relying upon this finding as their one source for determining the genders  from names. It was also for this reason that we have released as much of the data used in this study as possible in a repository (currently not cited due to author anonymity requirements). 
    \item {\bf Small-sample Accuracy Estimates:} Accuracy estimates of the various NLP steps noted in earlier sections were derived from fairly small samples and may be susceptible to  bias. Future researchers should not quote or trust those estimates blindly, but aim to do their own sampling and annotation to expand the annotated sample set, discover potential biases in our own sample set, and derive accuracy estimates with higher statistical power. 
    \item {\bf Possible Methodological Bias:} We advocate for more scrutiny into whether our methods, and the manner in which we investigated our hypotheses (or even the formulation of the hypotheses) might have been skewed or biased in a way that is not apparent to us at present. For instance, there is always the possibility that if the hypothesis had been stated a different way, or if we had used other measures of character prevalence, that the findings may have indicated a different degree of gender bias than what we reported. Another problem is that, as recent work as shown, there is considerable gender bias in seemingly unbiased computational systems, including NLP systems \cite{genderbiasNLP}. Therefore, we hope that future researchers will consider alternative ways of formulating gender-relevant hypotheses, deriving intermediate data structures, and replicating the study using the dataset we have made available. 
\end{enumerate}

\section{Conclusion}

In this article, we defined and measured the differences between male character prevalence and female character prevalence using three robust measures of prevalence, on a corpus of copyright-expired literary texts from the Project Gutenberg English-language corpus. Using computationally replicable methodologies relying on modern natural language processing tools, we found that female character prevalence is significantly lower than that of male character prevalence, although the difference declines (while still being significant) when controlling for the gender of the author. We also found that male character ratios have not varied much over time in our sample. Recent results, especially reported in \emph{Related Work}, are consistent with this finding. More broadly, we hope that our findings serve as a case study illustrating the promise of using open-source tools and data, in conjunction with careful quality control and statistical analysis, to  objectively explore gender-specific issues of socio-cultural inequality.


\bibliographystyle{ACM-Reference-Format}
\bibliography{sample-sigconf}










\end{document}